\documentclass{article}
\usepackage{iclr2027_conference,times}
\usepackage{graphicx}
\usepackage{hyperref}
\usepackage{url}
\renewcommand{\ttdefault}{cmtt}
\usepackage{amsmath}
\usepackage{booktabs}
\usepackage{array}
\usepackage{xcolor}
\usepackage{placeins}
\hypersetup{hidelinks}

\title{Reuse, Rewrite, or Reject: Evidence-Constrained Trace Reuse for Repeated Mobile GUI Control}
\author{
Bingyu Bai, Cheng Zhang, Erhu Feng%
\thanks{Corresponding author.} \\
Institute of Parallel and Distributed Systems (IPADS) \\
Shanghai Jiao Tong University \\
\texttt{\{ashley-bby,sjtuzc954,fengerhu1\}@sjtu.edu.cn}
}

\iclrfinalcopy

\begin{document}

\raggedbottom

\maketitle
\lhead{}

\begin{abstract}
Mobile GUI agents frequently perform repetitive tasks with similar interaction procedures. Reusing previous executions can reduce redundant LLM inference and execution overhead, but variations in user instructions and execution environments may invalidate previously successful actions. We systematically characterize these variations into six representative categories and observe that such changes do not necessarily invalidate entire procedures: individual transitions can remain reusable when validated against the current task context and interface state.

Based on this insight, we propose graph--text templates for selective execution-trace reuse. The graph captures UI states and action transitions, while the text encodes task intent, procedural structure, and typed slots for variable content. A record--template--replay framework constructs templates from verified executions and leverages the current instruction and live UI state to reuse, adapt, or discard individual transitions. Given an execution trace, our framework reduces LLM inference token consumption by approximately 80\% on average across five base agents on AndroidWorld, with most agents achieving at least 1.5$\times$ execution speedup over LLM-only execution. In real-device evaluations on Android and HarmonyOS, our framework achieves action reuse ratios of 70.7--79.8\% while maintaining over 2$\times$ execution speedup in most evaluated configurations.
\end{abstract}

\section{Introduction}

A central goal for mobile GUI agents is to become faster as they accumulate experience. In everyday use, an agent repeatedly encounters familiar tasks and interaction patterns, creating opportunities to reuse what it has already learned from execution. An agent that invokes a large language model (LLM) before every action continues to pay for decisions it has successfully made before. Turning previous executions into reusable experience can reduce these repeated model calls and user waiting time. The question is how to sustain this benefit as task conditions change.

Two obstacles remain: changes in the environment and instruction. Even when the request remains the same, the application may be in a different state, its interface may have changed, or a dialog may interrupt the expected workflow. Conversely, a familiar interface may receive a request with a different value, target, or wording. These changes can make a previously successful action unnecessary, inappropriate, or insufficient for the current task. Repeated use therefore requires an agent to determine which parts of its experience remain useful under the current conditions. Addressing these two sources of variation is central to making experience translate into faster execution.

We organize these obstacles into six main categories. Environment changes comprise background-state changes, layout and workflow changes, and modal interruptions; instruction changes comprise parameter changes, entity substitutions, and goal-expression changes. This categorization guides our analysis of when historical steps can be retained, need adaptation, or should be abandoned. We introduce their implications for reuse in Section~\ref{sec:method} and develop the detailed case analysis in Section~\ref{sec:experiments}.

MobiMem regenerates variable-step actions when parameters change, while SkillDroid and PreAct execute parameterized skills or programs \citep{liu2025mobimem,chen2026skilldroid,li2026preact}. We instead organize reuse around candidate transitions in a unified graph--text template: semantic descriptions retrieve executable structure, local annotations expose replaceable content, and live evidence determines whether each transition is reused, rewritten, or rejected. Each admitted transition is further guarded by pre- and post-action evidence checks, with rejected or failed steps falling back to the base agent.

\begin{figure}[t]
  \centering
  \includegraphics[page=1,width=0.98\textwidth]{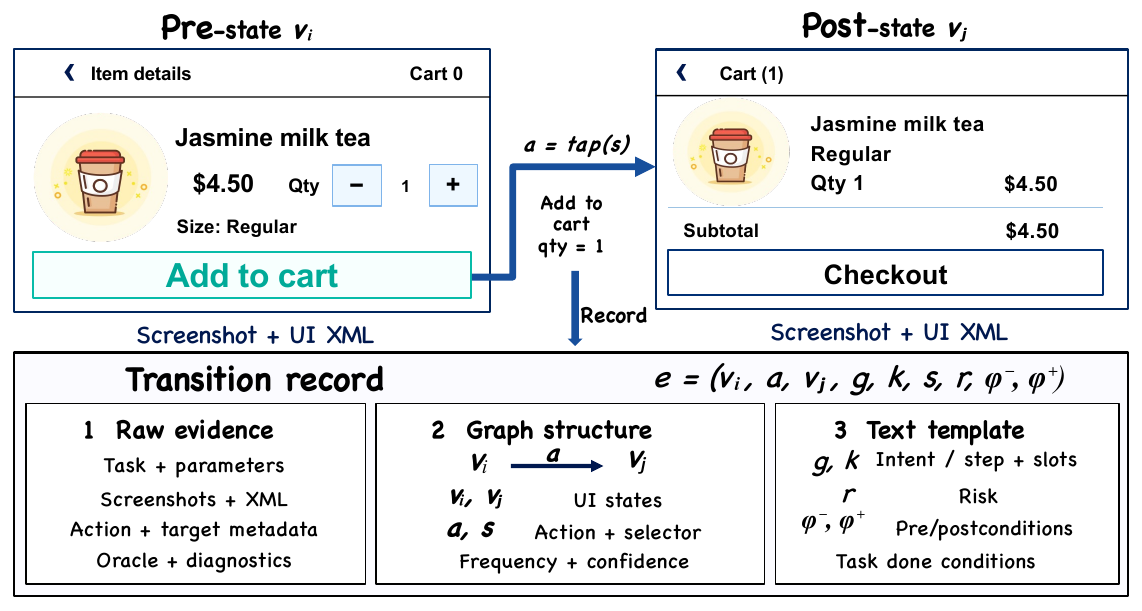}
  \caption{A recorded transition links an action to checked before-and-after UI evidence. The illustrative add-to-cart action connects the selected jasmine milk tea and quantity to the resulting cart state. The lower panels organize evidence into a graph and text template for reuse decisions.}
  \label{fig:motivation}
  \vspace{-18pt}
\end{figure}

We propose an evidence-constrained record--template--replay framework. During recording, the agent stores each action with its target and checked before-and-after states (Figure~\ref{fig:motivation}). The add-to-cart example links the selected item and quantity to the resulting cart state. Template extraction then preserves reusable procedure structure, exposes variable content as typed slots, and retains incompatible paths as separate branches. The resulting graph--text template integrates a capability description for semantic retrieval with executable UI transitions. Within the same template, local annotations mark entities and values that can be rebound to the current request, allowing applicable parts of a recorded procedure to be reused while changed steps are rewritten. During replay, the current instruction and live interface determine whether a candidate transition can be reused, rewritten with an explicitly supplied value, or rejected. Each admitted transition undergoes a pre-check, one-step execution, and a post-check; insufficient evidence or a failed check returns control to the same base agent. The controller thus reuses applicable steps while adapting to or recovering from changes elsewhere.

We first evaluate the framework on AndroidWorld, which provides reproducible mobile tasks with parameterized initialization and programmatic outcome checks \citep{rawles2024androidworld}. Across five base agents, we measure task success, executed reuse, token consumption, LLM calls, and execution time. Given a verified execution record, our framework achieves 82.0--87.5\% executed reuse and 1.2--1.9$\times$ execution-time speedup. These tests establish a basis for assessing repeated execution, while comparison with practical phone use highlights differences in devices, operating systems, interface languages, and operating conditions. Although AndroidWorld supports variation in task parameters and initial states, its benchmark setting alone cannot establish how reuse performs across these real-use conditions. This motivates extending the evaluation to physical devices.

We therefore conduct real-device evaluations on OnePlus 13, OPPO Find X8, Redmi K50, and Huawei Pura 80, spanning Android and HarmonyOS. Building on the six-category analysis, the quantitative studies focus on three representative conditions: background-state changes, parameter changes, and paraphrased instructions. These conditions examine whether the controller can account for a changed state, bind a newly requested value, and retrieve a familiar procedure under different wording. We report task success and execution efficiency, with component ablations examining state rechecking, value binding, and semantic retrieval. Across these conditions, our framework achieves 1.8--3.9$\times$ execution-time speedup. Together, these studies examine repeated-execution efficiency and the mechanisms needed to reuse experience under environment and instruction changes.

Our contributions are threefold. First, we analyze environment and instruction changes as two central obstacles to faster repeated GUI execution and organize them into six main categories of variation. Second, we develop graph--text transition templates and evidence-constrained selective replay to retain applicable actions, adapt explicitly bound values, and return control to the base agent when reuse is unsupported. Third, we combine AndroidWorld evaluation with experiments on four phone models spanning Android and HarmonyOS, measuring execution efficiency and task outcomes and using targeted ablations to examine adaptation to state, parameter, and wording changes.

\section{Related Work}

\textbf{GUI Agents for Mobile.}
AutoDroid and AppAgent support mobile automation through app knowledge or demonstrations, while UI-TARS integrates perception and action modeling \citep{wen2023autodroid,zhang2023appagent,qin2025uitars}. For efficient execution with smaller models, AutoDroid-V2 generates scripts from app documentation, Ferret-UI Lite develops compact policies, MobileExplorer overlaps inference with exploration, and UI-KOBE constrains runtime decisions through app graphs \citep{wen2025autodroidv2,yang2025ferretuilite,huang2026mobileexplorer,chai2026uikobe}. These approaches address model and planning costs through complementary mechanisms.

\textbf{Memory and Skill Learning for GUI Agents.}
Agent Workflow Memory, Agent S, and LearnAct organize workflows, experience, or demonstrations to support subsequent execution, with Agent Workflow Memory also offering an executable action-space variant \citep{wang2024awm,agashe2024agents,liu2026learnact}. UI-Mem, MementoGUI, and MMSkills further use hierarchical experience or multimodal memory and procedures \citep{xiao2026uimem,zeng2026mementogui,zhang2026mmskills}. Executable skill construction takes several forms: callable APIs in SkillWeaver, recursive parameterized skills in SCAFFOLD, and state-machine memory or execution graphs in ActionEngine and CUA-Skill \citep{zheng2025skillweaver,he2026scaffold,zhong2026actionengine,chen2026cuaskill}. \mbox{WebXSkill} and SGDR pair executable skills with text for context-aware retrieval \citep{wang2026webxskill,li2026sgdr}. EvoSkill-GUI, KnowAct-GUIClaw, and online skill evolution maintain and revise persistent procedures \citep{chen2026evoskill,li2026knowact,hu2026skillevolution}.

\textbf{Trace Replay for GUI Agents.}
V2S and GPA derive replayable procedures from recordings or demonstrations \citep{havranek2021v2s,zhao2026gpa}. Mobile-Agent-E and AppAgentX learn executable shortcuts, while AgentRR adds experience abstractions with precondition and parameter checks \citep{wang2025mobileagente,jiang2025appagentx,feng2025agentrr}. MobileGPT, MobiAgent, and MobiMem use structured memories to reuse subtasks or local actions: MobileGPT adapts parameters and UI attributes with model-assisted parameter filling, MobiAgent combines action caching with UI-change detection, and MobiMem reuses prefixes and suffixes around steps requiring fresh reasoning \citep{lee2024mobilegpt,zhang2025mobiagent,liu2025mobimem}. Executable Agentic Memory uses a lightweight value model for graph search, whereas MemoryPlayback uses confidence to switch between replay and VLM reasoning \citep{qin2026eam,luo2026memoryplayback}. PreAct, SkillDroid, and EchoPath combine parameterized replay with execution checks and agent fallback or repair \citep{li2026preact,chen2026skilldroid,zhao2026echopath}. SkillDroid checks state at each step and skips satisfied steps; its semantic matching invokes an LLM for confirmation and slot extraction. Selective replay, parameter adaptation, and fallback are therefore established capabilities. The comparison concerns admission/rewrite evidence under changed intent and state, and its acquisition cost.

\textbf{Action Verification for GUI Agents.}
VeriSafe Agent checks actions against formalized user intent, while SenseAct uses typed control primitives and programmatic postconditions \citep{lee2025verisafe,cai2026senseact}. AppAgent-Claw validates outcomes after execution \citep{song2026appagentclaw}; \mbox{ReUseIt}, ContractSkill, and SKILL.nb combine pre/postcondition checks with local recovery in reusable workflows or skills \citep{liu2026reuseit,lu2026contractskill,elhattami2026skillnb}. These approaches address action validity and outcome correctness, informing comparisons of pre-reuse checks and post-action verification.

\textbf{Benchmarks for GUI Agents.}
AndroidWorld and MobileWorld evaluate mobile task execution, while MemGUI-Bench targets memory-dependent tasks \citep{rawles2024androidworld,kong2025mobileworld,liu2026memguibench}. Final task success alone can obscure incorrect reused actions repaired by subsequent recovery, motivating analysis of local replay decisions alongside overall success and execution cost.

We study evidence-constrained transition reuse under changing instructions and UI states, using graph--text contracts to select reuse, rewriting, or rejection with fallback to the same base agent.

\section{Method}
\label{sec:method}
\edef\savedmethodtopnumber{\arabic{topnumber}}
\setcounter{topnumber}{1}

To make repeated mobile tasks faster, the controller must replace selected LLM calls without replaying actions that have become stale. Both the execution environment and instruction can change. We separate memory formation from memory use: the \emph{record} phase captures evidence from an execution, template extraction turns that evidence into a reusable graph--text procedure, and the \emph{replay} phase admits one transition at a time under live checks. The resulting decision is \textsc{reuse}, \textsc{rewrite}, or \textsc{reject}; rejection or a failed postcheck returns control to the fallback policy.

\begin{figure}[t]
  \centering
  \includegraphics[page=2,width=0.98\textwidth]{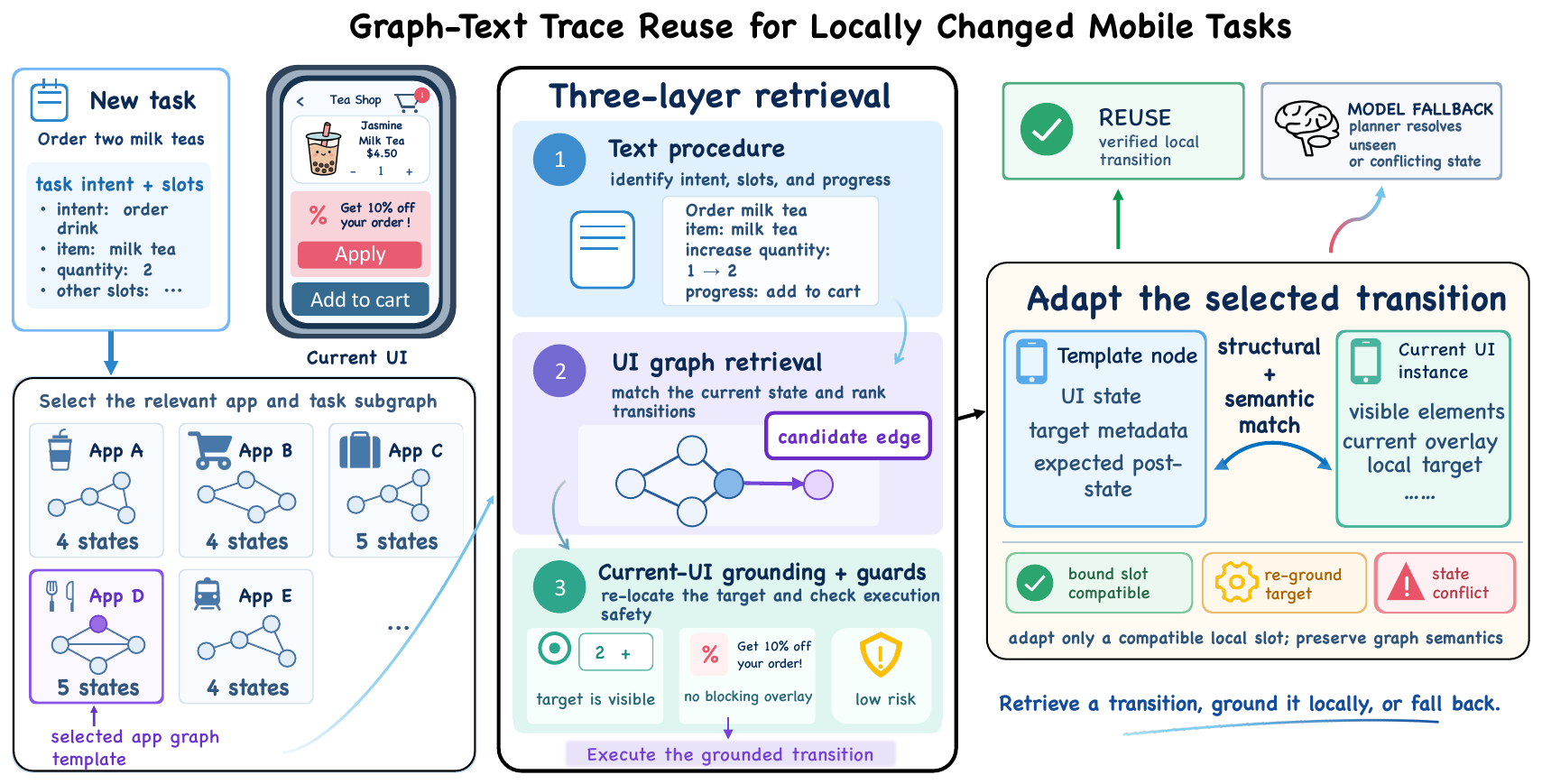}
  \caption{A changed request is matched to a candidate transition through task semantics, UI structure, and live grounding. In the illustrative milk-tea task, the current request supplies quantity 2. The controller retrieves a compatible app subgraph, binds the current slots, and checks the live target before admitting a local action or falling back to the model.}
  \label{fig:construction}
  \vspace{-18pt}
\end{figure}

\textbf{Environment and instruction changes.}
Historical actions can become invalid under an unchanged instruction. \emph{Background state and TOCTOU} may make a goal true at check time but false at action time, so duplicate or inverse actions are suppressed and the state is checked again before execution. \emph{Layout and workflow drift} may move a target or change the required path; semantic relocation can handle the former, while the latter requires fallback. \emph{Overlays and interruptions} can block the underlying page; only separately verified benign overlays are handled locally, whereas permission, login, payment, and task-relevant dialogs remain blocking states. These changes usually invalidate individual transitions rather than entire trajectories. The same environment can receive a different request. \emph{Parametric changes} in quantity, time, or amount can be rewritten only when the live instruction supplies a matching typed slot. \emph{Entity changes} require re-grounding products, contacts, or places in the current UI. \emph{Paraphrase and reference changes} can share a canonical procedure when their intent is supported; an under-specified preference reference is rejected rather than silently filled. Instruction changes allow constrained local rewriting, not implicit replacement of critical values.

\textbf{Recording transition evidence.}
During \emph{record}, the fallback agent completes the task normally while the system stores the evidence needed for later decisions: the observable state before and after each action, the action and its target, the procedure step, any typed value, and the resulting progress or completion signal. The paired screens in Figure~\ref{fig:motivation} illustrate how an action is associated with its observed effect. The unit of memory is a transition with a checked outcome, not an unexamined script. Failed terminal outcomes are not stored as successful demonstrations; a prefix is retained only when its truncation point and state evidence are explicit.

\textbf{Extracting reusable templates.}
Template extraction consolidates recorded transitions into a graph--text representation. Compatible observable states and actions are merged, while conflicting states remain separate branches. Values that may change across instructions are represented as typed slots; stable navigation and progress structure are kept as reusable edges. Transitions with missing evidence, an ungroundable target, or a failed terminal outcome are excluded from the released template. An ``ideal'' template is as reusable as the record supports without hard-coding incidental state.

The resulting transition contract is
\begin{equation}
e=(v_i,a,v_j,z), \quad z=(g,k,s,r,\phi^{-},\phi^{+}),
\end{equation}
where the text procedure supplies intent $g$ and typed slots $k$, the graph supplies source and target states $v_i,v_j$ and selectors $s$, and $\phi^{-},\phi^{+}$ encode the evidence required before and after the action. The text view states what progress is intended; the graph view states how that progress changed a concrete UI. These complementary views correspond to the graph and text panels in Figure~\ref{fig:motivation}.

\textbf{Guarded selective replay.}
During \emph{replay}, the current instruction and live UI select a candidate edge from the template. Figure~\ref{fig:construction} illustrates this process for a request to order two milk teas: the text procedure identifies the intent and quantity slot, UI-graph retrieval finds a compatible candidate, and live grounding checks its target and execution conditions. These retrieval checks identify a candidate; execution additionally requires evidence that its action remains valid for the current task.

\begin{figure}[t]
  \centering
  \includegraphics[page=3,width=0.98\textwidth]{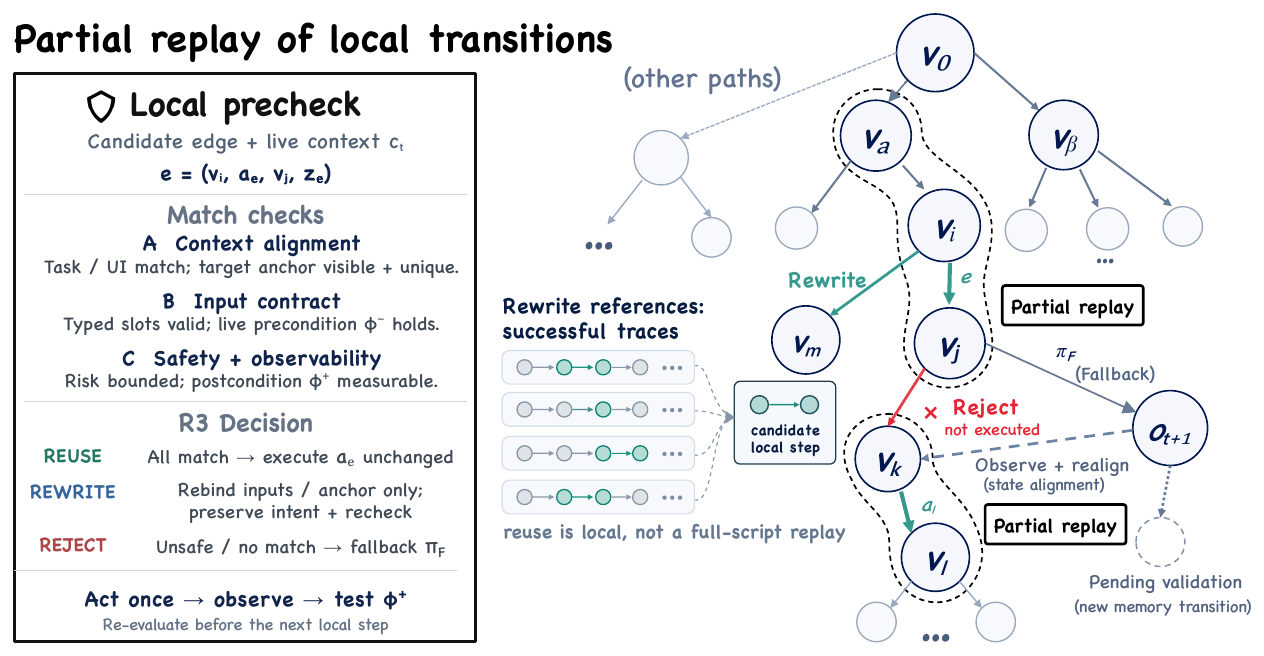}
  \caption{Replay proceeds through individually checked transitions. Left: live context and contract checks determine whether to reuse, rewrite, or reject a candidate. Right: the green branch illustrates local rewriting; the red rejected edge is not executed, and fallback provides a new observation for realignment. Admitted actions execute once and undergo a post-check before the next local decision.}
  \label{fig:method}
  \vspace{-18pt}
\end{figure}

Specifically, the candidate's procedure step must agree with the current instruction, and its source-state conditions and target selector must be supported by the live UI. Matching the historical page alone cannot justify copying its action payload. The controller emits \textsc{reuse} when the payload remains valid under these conditions, or \textsc{rewrite} when a changed value is explicitly bound to a compatible typed slot. If the required value or target cannot be established, it emits \textsc{reject} even when the navigation context matches. High-risk entity/value pairs require stronger evidence and cannot be filled from preference alone. Figure~\ref{fig:method} (left) summarizes the checks and decisions.

For an admitted edge, the controller refreshes the required evidence in a pre-check, executes the action once if the check passes, and performs a semantic post-check. The post-check must confirm the expected state change or completion predicate using the current slot bindings; a page change alone is not sufficient for every action. A failed post-check blocks the edge for the current episode and returns control to the fallback policy, preventing one mismatch from propagating through a trajectory. The right panel of Figure~\ref{fig:method} shows how a rejected edge remains unexecuted while fallback produces a new observation for realignment. Each subsequent transition is rechecked before execution.

For example, changing ``order one jasmine milk tea'' to ``order two jasmine milk teas'' connects the recorded transition in Figure~\ref{fig:motivation} to the live task in Figure~\ref{fig:construction}. Valid navigation transitions remain reusable, while the quantity-setting action is rebound to 2 and checked against the live UI. Its post-check must confirm the displayed quantity. If the quantity control cannot be grounded, that transition is rejected and control returns to the base agent, following Figure~\ref{fig:method}. Adding the item to the cart is a separate transition whose post-check must confirm the requested item and quantity; it does not itself establish completion of the order. Thus, both action and verification change.

\FloatBarrier
\setcounter{topnumber}{\savedmethodtopnumber}

\section{Experiments}
\label{sec:experiments}

We evaluate whether historical experience can support reliable and efficient repeated mobile execution. We report success and efficiency on AndroidWorld and on real phones under background-state changes, parameter changes, and paraphrased instructions. Three real-device ablations examine how to preserve valid actions, adapt values, and retrieve procedures across paraphrases.

\subsection{Overall Results on AndroidWorld and Real-World Tasks}

\paragraph{Benchmark and agents.}
AndroidWorld is a widely adopted benchmark for online mobile GUI agents. Its original suite contains 116 tasks across 20 Android applications, with parameterized task initialization and programmatic success evaluation \citep{rawles2024androidworld}. We evaluate our replay-assisted pipeline with GUI-Owl-7B, MAI-UI-8B, GELab-Zero-4B-preview, AutoGLM-Phone-9B-Multilingual, and UI-TARS-1.5-7B. For each base agent, successful LLM-only executions supply the records used to construct its template graphs. Record and replay tasks are similar, with some pairs differing in task details or parameter values to test reuse and rewriting; each candidate action is checked against the live instruction and interface, with the same base agent handling fallback. For context, PreAct reports 73.3\% warm-run success on a subset of 15 tasks selected from AndroidWorld~\citep{li2026preact}.

\paragraph{Metrics and evaluation scope.}
Table 1 covers AndroidWorld and three real-world conditions.
For our five base agents, Success rate equals base success
(LLM-only on AndroidWorld; replay-only in real-world blocks)
times replay-and-fallback success, conditional on successful
base-agent records for AndroidWorld, and thus cannot exceed base success.
The real-world ablations below retain per-configuration success rates.
Efficiency uses a separate LLM-only cost reference on matched live tasks;
record/live pairs may differ. In real-world blocks, this reference
differs from the replay-only success baseline. Efficiency comparisons
require both the LLM-only baseline and the full controller to succeed
on the same task instance with identical initial states and completion
criteria. Failed executions remain in the task-success denominator
but are excluded from efficiency comparisons.
Execution-time speedup is the arithmetic mean of per-task
baseline-to-controller time ratios.

\begin{table}[!ht]
  \vspace{-8pt}
  \centering
  \caption{AndroidWorld and real-world results for five base agents. Success, Replay successful rate, replay, token reduction, and LLM-step reduction are percentages (higher is better); execution-time speedup is a multiplier.}
  \label{tab:overall-results}
  \small
  \setlength{\tabcolsep}{2pt}
  \renewcommand{\arraystretch}{1.16}
  \begin{tabular*}{\linewidth}{@{\extracolsep{\fill}}cccccccc@{}}
  \toprule
  Model & \shortstack{Original\\success score} & \shortstack{Replay\\successful\\rate} & \shortstack{Success score\\with replay} & \shortstack{Replay\\rate} & \shortstack{Token\\$\downarrow$ (\%)} & \shortstack{LLM steps\\$\downarrow$ (\%)} & \shortstack{Execution\\speedup} \\
  \midrule
  \multicolumn{8}{c}{\textbf{AndroidWorld}} \\
  \midrule
  GUI-Owl-7B & 59.1 & 97.7 & 57.7 & 87.5 & 89.6 & 79.8 & 1.7$\times$ \\
  MAI-UI-8B & 66.4 & 95.5 & 63.4 & 86.4 & 78.3 & 72.5 & 1.9$\times$ \\
  GELab-Zero-4B & 71.7 & 94.5 & 67.8 & 85.6 & 78.0 & 75.8 & 1.5$\times$ \\
  AutoGLM-Phone-9B & 51.0 & 95.9 & 48.9 & 83.4 & 89.0 & 76.4 & 1.4$\times$ \\
  UI-TARS-1.5-7B & 21.0 & 94.0 & 19.7 & 82.0 & 79.2 & 71.1 & 1.2$\times$ \\
  \midrule
  \multicolumn{8}{c}{\textbf{Real-world: background-state changes}} \\
  \midrule
  GUI-Owl-7B & 51.0 & 96.0 & 49.0 & 78.0 & 65.4 & 88.8 & 1.9$\times$ \\
  MAI-UI-8B & 55.0 & 96.5 & 53.1 & 75.8 & 72.3 & 84.5 & 2.5$\times$ \\
  GELab-Zero-4B & 60.0 & 98.0 & 58.8 & 72.6 & 78.9 & 80.7 & 3.6$\times$ \\
  AutoGLM-Phone-9B & 52.0 & 96.0 & 49.9 & 75.8 & 78.4 & 88.4 & 3.6$\times$ \\
  UI-TARS-1.5-7B & 49.0 & 93.0 & 45.6 & 72.9 & 71.8 & 80.6 & 2.7$\times$ \\
  \midrule
  \multicolumn{8}{c}{\textbf{Real-world: parameter changes}} \\
  \midrule
  GUI-Owl-7B & 46.0 & 96.0 & 44.2 & 74.9 & 78.7 & 75.5 & 2.2$\times$ \\
  MAI-UI-8B & 53.0 & 95.0 & 50.4 & 70.7 & 68.3 & 80.6 & 3.5$\times$ \\
  GELab-Zero-4B & 54.0 & 98.0 & 52.9 & 79.8 & 73.9 & 85.8 & 2.1$\times$ \\
  AutoGLM-Phone-9B & 48.0 & 94.0 & 45.1 & 76.4 & 69.1 & 80.3 & 2.0$\times$ \\
  UI-TARS-1.5-7B & 44.0 & 93.0 & 40.9 & 73.8 & 72.4 & 73.1 & 1.8$\times$ \\
  \midrule
  \multicolumn{8}{c}{\textbf{Real-world: paraphrased instructions}} \\
  \midrule
  GUI-Owl-7B & 55.0 & 94.0 & 51.7 & 79.2 & 62.5 & 85.9 & 2.9$\times$ \\
  MAI-UI-8B & 52.0 & 96.0 & 49.9 & 73.8 & 68.2 & 86.8 & 3.6$\times$ \\
  GELab-Zero-4B & 57.0 & 92.0 & 52.4 & 75.0 & 79.6 & 85.0 & 3.9$\times$ \\
  AutoGLM-Phone-9B & 51.0 & 97.0 & 49.5 & 76.4 & 67.3 & 83.5 & 3.6$\times$ \\
  UI-TARS-1.5-7B & 49.0 & 93.0 & 45.6 & 73.2 & 60.8 & 76.0 & 3.6$\times$ \\
  \bottomrule
  \end{tabular*}
  \par\smallskip
  \begin{minipage}{\linewidth}
  Base success denotes LLM-only execution in AndroidWorld and replay-only execution in the real-world blocks.
  \end{minipage}
  \par\vspace{-14pt}
\end{table}

\paragraph{Replay-rate definitions.}
For episode $i$, let $R_i$ be the number of executed actions supplied by replay without an LLM call, and $N_i$ the total number of executed actions, including replay, fallback, and recovery. Locally rewritten actions count as replay only when no model call is needed; retrieved but unexecuted candidates do not count. Two useful summaries are
\begin{equation}
  r_{\mathrm{task}}=\frac{1}{M}\sum_{i=1}^{M}\frac{R_i}{N_i},
  \qquad
  r_{\mathrm{step}}=\frac{\sum_i R_i}{\sum_i N_i},
  \label{eq:replay-rates}
\end{equation}
where $M$ counts episodes with at least one executed action. The first gives each task equal weight; the second gives each executed step equal weight.

\paragraph{From replay to fewer calls, tokens, and lower latency.}
Let $N_0$ and $N_1$ be the pooled GUI-action counts without and with replay on paired tasks. With one model call per baseline action and per non-replayed assisted action, and $C_{\mathrm{extra}}$ additional assisted calls for model-based checking or recovery, the call counts are $C_0=N_0$ and $C_1=(1-r_{\mathrm{step}})N_1+C_{\mathrm{extra}}$. The reductions are
\begin{equation}
  d_{\mathrm{LLM}}=1-\frac{C_1}{C_0}
  =1-\frac{(1-r_{\mathrm{step}})N_1+C_{\mathrm{extra}}}{N_0},
  \qquad
  d_{\mathrm{tok}}=1-\frac{K_1}{K_0},
  \label{eq:call-token-savings}
\end{equation}
where $K_0$ and $K_1$ include all billed input and output tokens, including visual tokens and any model-based checking or recovery. The table expresses reductions as percentages. LLM-step reduction counts avoided model decisions; physical GUI-action reduction is a distinct quantity, $1-N_1/N_0$, because replay still executes actions on the device. Token reduction also depends on the input and output lengths of the remaining calls; neither reduction is estimated from replay rate in Table~\ref{tab:overall-results}.

Execution-time speedup is $S=T_0/T_1$, using matched timing boundaries. To explain its difference from the fraction of avoided LLM calls, let $\tau$ be the average model-call latency and $\epsilon$ the local replay overhead per replayed action. With equal call costs and no extra calls, the decision-time ratio is
\begin{equation}
  S_{\mathrm{decision}}
  =\frac{N_0\tau}{(N_1-R)\tau+R\epsilon}
  \underset{\epsilon/\tau\to 0}{\approx}
  \frac{1}{(1-r_{\mathrm{step}})N_1/N_0}
  =\frac{1}{1-d_{\mathrm{LLM}}}\quad(C_{\mathrm{extra}}=0),
  \label{eq:ideal-speedup}
\end{equation}
where $R=\sum_i R_i$. This limit covers only model decisions and local replay. Execution-time speedups also include UI-structure acquisition, device actions, waits, retrieval, verification, and recovery; they do not follow this idealization. Recording and template construction costs are separate.

\paragraph{Results.}
Table~\ref{tab:overall-results} reports 19.7--67.8\% success on AndroidWorld, with 82.0--87.5\% executed reuse and 1.2--1.9$\times$ execution-time speedup. Across the three real-device conditions, the computed success rates range from 40.9\% to 58.8\%, with 1.8--3.9$\times$ speedup. Token consumption and LLM calls decrease across all reported settings.

\subsection{Real-World Evaluation and Ablations}

\textbf{Real-device evaluation is essential} for understanding how reduced LLM inference translates into practical execution speedups across devices and operating systems. Our timing measurements on AndroidWorld show that pre-action preparation and post-action state acquisition together can even account for 40.02\% of total execution time. The benchmark protocol involves repeated UI-structure acquisition during these stages, reducing the fraction of runtime spent on LLM inference and thereby limiting the overall speedup achievable by avoiding model calls. This helps explain the comparatively modest speedups on AndroidWorld despite substantial reductions in LLM calls. We therefore extend our evaluation to four phone models spanning Android and HarmonyOS, optimizing UI-structure acquisition and reducing its frequency to lower observation overhead. This increases the fraction of runtime that can benefit from trace reuse, allowing us to assess its acceleration benefits alongside robustness to changes in device state and user instructions.

Table~\ref{tab:taxonomy} summarizes six categories of variation in real-world phone use. We evaluate three representative conditions: background-state changes preserve the instruction while changing the device state; parameter changes modify a requested value within a familiar procedure; and paraphrased instructions change wording while preserving intent.

\begin{table}[htbp]
\vspace{-8pt}
\centering
\small
\caption{Two sources and six categories of variation in real-world phone use. Ablations examine background-state and parameter changes, and paraphrases within goal-expression changes.}
\label{tab:taxonomy}
\setlength{\tabcolsep}{4pt}
\renewcommand{\arraystretch}{1.12}
\begin{tabular}{>{\centering\arraybackslash}m{\dimexpr0.20\linewidth-2\tabcolsep\relax}@{}c}
\toprule
Change source &
\begin{tabular}[c]{>{\centering\arraybackslash}m{\dimexpr0.28\linewidth-2\tabcolsep\relax}>{\centering\arraybackslash}m{\dimexpr0.52\linewidth-2\tabcolsep\relax}}
Category & Implication for trace reuse
\end{tabular} \\
\midrule
\shortstack[c]{Environment\\changes} &
\begin{tabular}[c]{>{\centering\arraybackslash}m{\dimexpr0.28\linewidth-2\tabcolsep\relax}>{\centering\arraybackslash}m{\dimexpr0.52\linewidth-2\tabcolsep\relax}}
Background-state changes & Check the live state and preserve satisfied goals; avoid duplicate or inverse actions. \\
Layout and workflow changes & Re-ground displaced targets and reject paths invalidated by a changed workflow. \\
Modal interruptions & Detect blocking overlays and reassess which actions remain executable.
\end{tabular} \\
\midrule
\shortstack[c]{Instruction\\changes} &
\begin{tabular}[c]{>{\centering\arraybackslash}m{\dimexpr0.28\linewidth-2\tabcolsep\relax}>{\centering\arraybackslash}m{\dimexpr0.52\linewidth-2\tabcolsep\relax}}
Parameter changes & Bind current quantities, times, or other values and verify the resulting state. \\
Entity substitutions & Resolve the requested product, contact, or destination in the current interface. \\
Goal-expression changes & Retrieve procedures across paraphrases; require supporting evidence for contextual references.
\end{tabular} \\
\bottomrule
\end{tabular}
\vspace{-10pt}
\end{table}

\begin{figure*}[!ht]
  \vspace{-8pt}
  \centering
  \includegraphics[width=\linewidth]{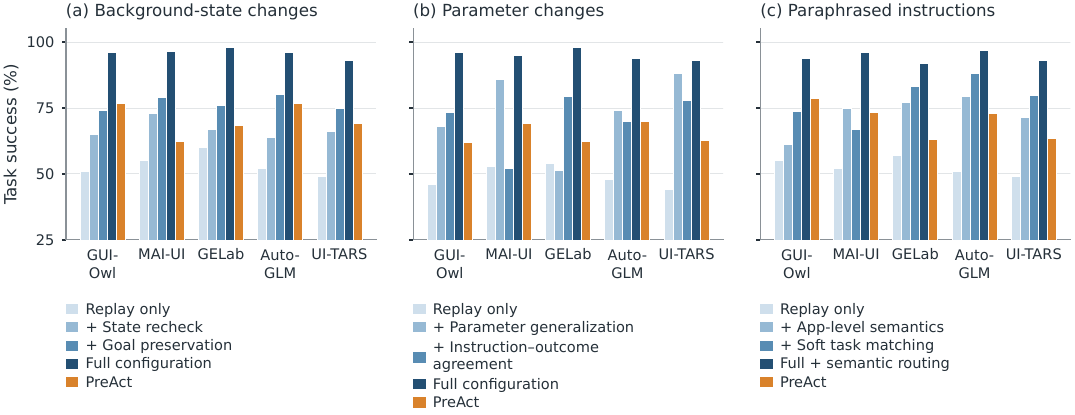}
  \caption{The full controller improves task success over replay only across all five agents under background-state changes, parameter changes, and paraphrased instructions. Each model group has four configurations in blue shades darkening from left to right, followed by PreAct in orange, with matching legend swatches below each panel. The vertical scale starts at 25\%. Component additions are cumulative across the blue bars. Model abbreviations refer to the five agents in Table~\ref{tab:overall-results}.}
  \label{fig:real-device-ablations}
  \vspace{-18pt}
\end{figure*}

\paragraph{Evaluation protocol.}
We evaluate record--live task pairs on physical Android and HarmonyOS phones. Each case compares replay only, two cumulative component additions, and the full controller under similar live instructions, initial states, and completion criteria. Figure~\ref{fig:real-device-ablations} reports human-assessed outcomes. For automated re-evaluation, we specify an independent task-specific verifier that checks the requested object, values, and terminal state against predefined predicates using accessible system/application state or structured UI evidence, following AndroidWorld's state-based evaluation principle~\citep{rawles2024androidworld}.

\paragraph{Ablation analysis.}
For background-state changes, parameter changes, and
paraphrased instructions, the full configuration improves success across all five agents; the full configuration reaches 93.0--98.0\% success, exceeding replay only by 38.0--45.0 percentage points. For parameter changes, parameter generalization and instruction--outcome agreement support value binding and verification; the full configuration achieves 93.0--98.0\% success, a gain of 42.0--50.0 percentage points. For paraphrased instructions, app-level semantics and soft task matching support procedure retrieval; the full configuration reaches 92.0--97.0\% success, a gain of 35.0--46.0 percentage points over replay only.

\paragraph{Comparison with Closely Related Methods.}
Direct comparisons with closely related reuse systems are constrained by implementation availability and differences in evaluation protocols. At the time of our artifact review, we could not locate publicly accessible official implementations of SkillDroid, EchoPath, or EAM \citep{chen2026skilldroid,zhao2026echopath,qin2026eam}. For MobiMem, MobiAgent, and AgentRR, we have not established a complete evaluation pipeline matching our record--live protocol from the available artifacts \citep{liu2025mobimem,zhang2025mobiagent,feng2025agentrr}. PreAct's controlled AndroidWorld replay experiments focus on a 15-task subset, limiting coverage and making the reported results potentially sensitive to task selection \citep{li2026preact}. MobileGPT's custom benchmark comprises 80 high-level tasks across eight apps, and its public implementation omits Random Explorer and human-in-the-loop memory repair, although the latter was also disabled in its original comparative evaluation \citep{lee2024mobilegpt}. Accordingly, we present externally reported results as contextual references and use within-system ablations to examine component contributions under our evaluation protocol.

\subsection{Discussion}

Given verified execution records, our results support selective trace reuse as a way to reduce repeated model inference and execution time in the evaluated settings. The framework combines applicable transitions, local rewriting of explicitly bound values, and fallback to the base agent within the same execution. This allows useful parts of prior experience to remain available as instructions and UI states change, with each reused action checked against current conditions.

\section{Limitations and Broader Impact}

Reuse depends on fallback coverage and UI observability; novel free-form content, inaccessible metadata, rapid visual changes, and unsupported workflows limit reuse. We evaluate five AndroidWorld base agents and selected real-device changes in background state, parameters, and instruction wording. Combined variations, cross-system trajectory transfer, and long-term adaptation need further study. Execution-time ratios omit recording and template-construction cost amortization.

Reusing private actions can expose stale or sensitive choices. Single-step admission, explicit high-risk slot checks, post-state validation, fallback, audit logs, and permission boundaries mitigate but do not eliminate this risk. Real-device studies should use reversible tasks or sandbox accounts; learned preferences must not be treated as authorization.

\section{Conclusion}

Experience can accelerate repeated mobile GUI execution when reusable procedure structure is preserved and each action is reassessed under current conditions. We analyzed six categories of environment and instruction changes and developed a record--template--replay framework that captures verified executions as graph--text templates. By combining task intent with live UI evidence, the controller retains applicable transitions, rewrites explicitly specified values, and invokes the base agent when reuse is unsupported. Given a verified execution record, evaluation across five base agents yields 1.2--1.9$\times$ execution-time speedup on AndroidWorld. In real-device evaluations on Android and HarmonyOS, our framework achieves action reuse ratios of 70.7--79.8\% while maintaining 1.8--3.9$\times$ execution-time speedup. Together, these findings support selective reuse as a means of translating past execution into fewer repeated model decisions and faster execution under changing task conditions.

\section*{Reproducibility Statement}
Section~3 describes the record--template--replay method, and Appendix~\ref{app:implementation} provides the transition schema and control-flow pseudocode. Section~\ref{sec:experiments} and Appendices~\ref{app:experimental-setup} and~\ref{app:metrics-results} describe the evaluation protocols, metric definitions, and aggregate results. Appendix~\ref{app:variation-cases} lists tested applications, presents neutralized task descriptions, and explains shared-suffix reuse and illustrative controlled variations. Appendix~\ref{app:reproducibility} describes the presentation and anonymization of task examples and privacy considerations for trajectory data.

\section*{AI Use Statement}
Generative AI tools assisted with manuscript drafting and language editing,
literature organization, and figure preparation. The authors reviewed the
resulting content and take full responsibility for the final manuscript.

\bibliography{references}

\begin{thebibliography}{44}
\providecommand{\natexlab}[1]{#1}
\providecommand{\url}[1]{\texttt{#1}}
\expandafter\ifx\csname urlstyle\endcsname\relax
  \providecommand{\doi}[1]{doi: #1}\else
  \providecommand{\doi}{doi: \begingroup \urlstyle{rm}\Url}\fi

\bibitem[Agashe et~al.(2025)Agashe, Han, Gan, Yang, Li, and
  Wang]{agashe2024agents}
Saaket Agashe, Jiuzhou Han, Shuyu Gan, Jiachen Yang, Ang Li, and Xin~Eric Wang.
\newblock {Agent S: An Open Agentic Framework that Uses Computers Like a
  Human}.
\newblock In \emph{International Conference on Learning Representations}, 2025.
\newblock URL
  \url{https://proceedings.iclr.cc/paper_files/paper/2025/hash/394c7c30ea87b5c3521b4d9e9d419071-Abstract-Conference.html}.

\bibitem[Cai et~al.(2026)Cai, Ma, Shao, Tang, Tsang, Lyu, and
  Yin]{cai2026senseact}
Hongtian Cai, Tianyi Ma, Jie-Jing Shao, Tianyi Tang, Ivor Tsang, Yueming Lyu,
  and Haiyan Yin.
\newblock {SenseAct: Structuring GUI Actions for Reliable Planning and
  Verification}.
\newblock In \emph{ICLR 2026 Workshop on Agents in the Wild}, 2026.
\newblock URL \url{https://openreview.net/attachment?id=0DznNQFW4g&name=pdf}.

\bibitem[Chai et~al.(2026)Chai, Xiao, Fu, Chen, Liu, and Li]{chai2026uikobe}
Yuxiang Chai, Han Xiao, Xinyu Fu, Jinpeng Chen, Rui Liu, and Hongsheng Li.
\newblock {UI-KOBE: Knowledge-Oriented Behavior Exploration for Lightweight
  Graph-Guided GUI Agents}, 2026.
\newblock URL \url{https://arxiv.org/html/2605.29534v1}.

\bibitem[Chen et~al.(2026{\natexlab{a}})Chen, Zhang, Tang, Lu, Du, Chen, Lu,
  Xiao, Zhuang, and Shen]{chen2026evoskill}
Bofan Chen, Boxuan Zhang, Fei Tang, Zhengxi Lu, Yong Du, Tongbo Chen, Weiming
  Lu, Jun Xiao, Yueting Zhuang, and Yongliang Shen.
\newblock {Reflect, Revise, Reuse: Training-Free Skill Evolution for GUI
  Agents}, 2026{\natexlab{a}}.
\newblock URL \url{https://arxiv.org/html/2609.17653v1}.

\bibitem[Chen et~al.(2026{\natexlab{b}})Chen, Bellucci, Sun, and
  Jacucci]{chen2026skilldroid}
Qijia Chen, Andrea Bellucci, Zhida Sun, and Giulio Jacucci.
\newblock {SkillDroid: Compile Once, Reuse Forever}, 2026{\natexlab{b}}.
\newblock URL \url{https://arxiv.org/html/2604.14872v1}.

\bibitem[Chen et~al.(2026{\natexlab{c}})Chen, Li, Solodko, Wang, Jiang, Cui,
  Hao, Ko, Abdali, Xu, Zheng, Fan, Cameron, Wagle, and
  Koishida]{chen2026cuaskill}
Tianyi Chen, Yinheng Li, Michael Solodko, Sen Wang, Nan Jiang, Tingyuan Cui,
  Junheng Hao, Jongwoo Ko, Sara Abdali, Leon Xu, Suzhen Zheng, Hao Fan,
  Pashmina Cameron, Justin Wagle, and Kazuhito Koishida.
\newblock {CUA-Skill: Develop Skills for Computer Using Agent},
  2026{\natexlab{c}}.
\newblock URL \url{https://arxiv.org/html/2601.21123v2}.

\bibitem[El~Hattami et~al.(2026)El~Hattami, Chapados, and
  Pal]{elhattami2026skillnb}
Amine El~Hattami, Nicolas Chapados, and Christopher Pal.
\newblock {SKILL.nb: Selective Formalization and Gated Execution for Durable
  Agent Workflows}, 2026.
\newblock URL \url{https://arxiv.org/html/2606.08049v1}.

\bibitem[Feng et~al.(2025)Feng, Zhou, Liu, Chen, Dong, Zhang, Zhao, Du, Hua,
  Xia, and Chen]{feng2025agentrr}
Erhu Feng, Wenbo Zhou, Zibin Liu, Le~Chen, Yunpeng Dong, Cheng Zhang, Yisheng
  Zhao, Dong Du, Zhichao Hua, Yubin Xia, and Haibo Chen.
\newblock {Get Experience from Practice: LLM Agents with Record \& Replay},
  2025.
\newblock URL \url{https://arxiv.org/html/2505.17716v1}.

\bibitem[Havranek et~al.(2021)Havranek, Bernal-C{\'a}rdenas, Cooper, Chaparro,
  Poshyvanyk, and Moran]{havranek2021v2s}
Madeleine Havranek, Carlos Bernal-C{\'a}rdenas, Nathan Cooper, Oscar Chaparro,
  Denys Poshyvanyk, and Kevin Moran.
\newblock {V2S: A Tool for Translating Video Recordings of Mobile App Usages
  into Replayable Scenarios}.
\newblock In \emph{IEEE/ACM 43rd International Conference on Software
  Engineering: Companion Proceedings}, 2021.
\newblock URL \url{https://arxiv.org/html/2103.04531v1}.

\bibitem[He et~al.(2026)He, Zhang, Ding, and Liu]{he2026scaffold}
Bowei He, Xiaokun Zhang, Meng Ding, and Xue Liu.
\newblock {SCAFFOLD: Self-Improving Web Agents via Recursive Parametric Skill
  Abstraction}, 2026.
\newblock URL \url{https://arxiv.org/html/2609.05511v1}.
\newblock Accepted to EMNLP 2026.

\bibitem[Hu et~al.(2026)Hu, Liang, and Zhu]{hu2026skillevolution}
Longtao Hu, Xiao Liang, and Linchao Zhu.
\newblock {From Interaction Traces to Persistent Skills: Online Evolution for
  Computer-Use Agents}, 2026.
\newblock URL \url{https://arxiv.org/html/2609.04869v1}.

\bibitem[Huang et~al.(2026)Huang, Zhang, Liu, and
  Ouyang]{huang2026mobileexplorer}
Runxi Huang, Liyu Zhang, Shengzhong Liu, and Xiaomin Ouyang.
\newblock {MobileExplorer: Accelerating On-Device Inference for Mobile GUI
  Agents via Online Exploration}, 2026.
\newblock URL \url{https://arxiv.org/html/2605.26546v1}.

\bibitem[Jiang et~al.(2025)Jiang, Zhuang, Song, Yang, Zhou, and
  Zhang]{jiang2025appagentx}
Wenjia Jiang, Yangyang Zhuang, Chenxi Song, Xu~Yang, Joey~Tianyi Zhou, and Chi
  Zhang.
\newblock {AppAgentX: Evolving GUI Agents as Proficient Smartphone Users},
  2025.
\newblock URL \url{https://arxiv.org/html/2503.02268v3}.

\bibitem[Kong et~al.(2025)Kong, Zhang, Yang, Gao, Liu, Tong, Cai, Zhou, Zhang,
  Chen, Liu, Hoi, and Wang]{kong2025mobileworld}
Quyu Kong, Xu~Zhang, Zhenyu Yang, Nolan Gao, Chen Liu, Panrong Tong, Chenglin
  Cai, Hanzhang Zhou, Jianan Zhang, Liangyu Chen, Zhidan Liu, Steven Hoi, and
  Yue Wang.
\newblock {MobileWorld: Benchmarking Autonomous Mobile Agents in Agent-User
  Interactive and MCP-Augmented Environments}, 2025.
\newblock URL \url{https://arxiv.org/html/2512.19432v3}.

\bibitem[Lee et~al.(2025)Lee, Lee, Choi, Im, Wi, Heo, Oh, Lee, and
  Shin]{lee2025verisafe}
Jungjae Lee, Dongjae Lee, Chihun Choi, Youngmin Im, Jaeyoung Wi, Kihong Heo,
  Sangeun Oh, Sunjae Lee, and Insik Shin.
\newblock {VeriSafe Agent: Safeguarding Mobile GUI Agent via Logic-based Action
  Verification}.
\newblock In \emph{Proceedings of the 31st Annual International Conference on
  Mobile Computing and Networking}, 2025.
\newblock \doi{10.1145/3680207.3765248}.
\newblock URL \url{https://prosys.kaist.ac.kr/publications/mobicom25.pdf}.

\bibitem[Lee et~al.(2024)Lee, Choi, Lee, Wasi, Choi, Ko, Oh, and
  Shin]{lee2024mobilegpt}
Sunjae Lee, Junyoung Choi, Jungjae Lee, Munim~Hasan Wasi, Hojun Choi, Steven~Y.
  Ko, Sangeun Oh, and Insik Shin.
\newblock {MobileGPT: Augmenting LLM with Human-like App Memory for Mobile Task
  Automation}.
\newblock In \emph{Proceedings of the 30th Annual International Conference on
  Mobile Computing and Networking}, 2024.
\newblock \doi{10.1145/3636534.3690682}.
\newblock URL \url{https://arxiv.org/html/2312.03003v3}.

\bibitem[Li(2026)]{li2026preact}
Bojie Li.
\newblock {PreAct: Computer-Using Agents that Get Faster on Repeated Tasks},
  2026.
\newblock URL \url{https://arxiv.org/html/2606.17929v1}.

\bibitem[Li et~al.(2026{\natexlab{a}})Li, Deng, Wang, Huang, Shi, Tan, Lu, and
  Liu]{li2026sgdr}
Jiaxi Li, Ke~Deng, Yun Wang, Jingyuan Huang, Yucheng Shi, Qiaoyu Tan, Jin Lu,
  and Ninghao Liu.
\newblock {Online Skill Learning for Web Agents via State-Grounded Dynamic
  Retrieval}, 2026{\natexlab{a}}.
\newblock URL \url{https://arxiv.org/html/2606.04391v1}.

\bibitem[Li et~al.(2026{\natexlab{b}})Li, Li, Su, Xu, Zhao, Bian, Liang, Zhang,
  Hu, and Zhang]{li2026knowact}
Yunxin Li, Jinchao Li, Shibo Su, Zhenran Xu, Chenrui Zhao, Tongshu Bian,
  Xiaoman Liang, Meishan Zhang, Baotian Hu, and Min Zhang.
\newblock {KnowAct-GUIClaw: Know Deeply, Act Perfectly, Personal GUI Assistant
  with Self-Evolving Memory and Skill}, 2026{\natexlab{b}}.
\newblock URL \url{https://arxiv.org/html/2607.12625v2}.

\bibitem[Liu et~al.(2026{\natexlab{a}})Liu, Zhao, Liang, Luo, Tang, Chai, Lin,
  Xiao, Wang, Chen, Lu, Wu, Wang, Liu, and Liu]{liu2026memguibench}
Guangyi Liu, Pengxiang Zhao, Yaozhen Liang, Qinyi Luo, Shunye Tang, Yuxiang
  Chai, Weifeng Lin, Han Xiao, WenHao Wang, Siheng Chen, Zhengxi Lu, Gao Wu,
  Hao Wang, Liang Liu, and Yong Liu.
\newblock {MemGUI-Bench: Benchmarking Memory of Mobile GUI Agents in Dynamic
  Environments}, 2026{\natexlab{a}}.
\newblock URL \url{https://arxiv.org/html/2602.06075v3}.
\newblock Accepted to ACM Multimedia 2026.

\bibitem[Liu et~al.(2026{\natexlab{b}})Liu, Zhao, Liu, Chen, Chai, Liang, Wang,
  Chen, Lu, Ren, Wang, He, Liu, and Meng]{liu2026learnact}
Guangyi Liu, Pengxiang Zhao, Liang Liu, Zhiming Chen, Yuxiang Chai, Yaozhen
  Liang, WenHao Wang, Siheng Chen, Zhengxi Lu, Shuai Ren, Hao Wang, Shibo He,
  Yong Liu, and Wenchao Meng.
\newblock {LearnAct: Few-Shot Mobile GUI Agent with a Unified Demonstration
  Benchmark}.
\newblock In \emph{Findings of the Association for Computational Linguistics:
  ACL 2026}, pp.\  29820--29843, 2026{\natexlab{b}}.
\newblock \doi{10.18653/v1/2026.findings-acl.1491}.
\newblock URL \url{https://aclanthology.org/2026.findings-acl.1491/}.

\bibitem[Liu et~al.(2026{\natexlab{c}})Liu, Sra, Inala, and
  Wang]{liu2026reuseit}
Yimeng Liu, Misha Sra, Jeevana~Priya Inala, and Chenglong Wang.
\newblock {ReUseIt: Synthesizing Reusable AI Agent Workflows for Web
  Automation}.
\newblock In \emph{Proceedings of the 31st International Conference on
  Intelligent User Interfaces}, pp.\  885--908. ACM, 2026{\natexlab{c}}.
\newblock \doi{10.1145/3742413.3789083}.
\newblock URL \url{https://arxiv.org/html/2510.14308v2}.

\bibitem[Liu et~al.(2025)Liu, Zhang, Zhao, Feng, Bai, Feng, Feng, Xia, and
  Chen]{liu2025mobimem}
Zibin Liu, Cheng Zhang, Xi~Zhao, Yunfei Feng, Bingyu Bai, Dahu Feng, Erhu Feng,
  Yubin Xia, and Haibo Chen.
\newblock {Beyond Training: Enabling Self-Evolution of Agents with MOBIMEM},
  2025.
\newblock URL \url{https://arxiv.org/html/2512.15784v1}.

\bibitem[Lu et~al.(2026)Lu, Zuo, Nie, He, Fan, Qi, and
  Jin]{lu2026contractskill}
Zijian Lu, Yiping Zuo, Yupeng Nie, Xin He, Weibei Fan, Lianyong Qi, and Shi
  Jin.
\newblock {ContractSkill: Repairable Contract-Based Skills for Multimodal Web
  Agents}, 2026.
\newblock URL \url{https://arxiv.org/html/2603.20340v3}.

\bibitem[Luo et~al.(2027)Luo, Tang, Han, Wang, Dong, Liu, Liang, He, Wang, and
  Zhang]{luo2026memoryplayback}
Qinyi Luo, Shunye Tang, Rongduo Han, Meiping Wang, Siyi Dong, Hong Liu, Xiao
  Liang, Chenghao He, Wei Wang, and Haining Zhang.
\newblock {MemoryPlayback: Leveraging Trajectory-Derived Priors for Robust
  Mobile GUI Execution}.
\newblock In \emph{Advanced Intelligent Computing Technology and Applications
  (ICIC 2026)}, volume 3040 of \emph{Communications in Computer and Information
  Science}, pp.\  85--96. Springer, Singapore, 2027.
\newblock \doi{10.1007/978-981-92-3557-5_7}.
\newblock URL
  \url{https://link.springer.com/chapter/10.1007/978-981-92-3557-5_7}.

\bibitem[Qin et~al.(2025)Qin, Ye, Fang, Wang, Liang, Tian, Zhang, Li, Li,
  Huang, Zhong, Li, Yang, Miao, Lin, Liu, Jiang, Ma, Li, Xiao, Cai, Li, Zheng,
  Jin, Li, Zhou, Wang, Chen, Li, Yang, Liu, Lin, Peng, Liu, and
  Shi]{qin2025uitars}
Yujia Qin, Yining Ye, Junjie Fang, Haoming Wang, Shihao Liang, Shizuo Tian,
  Junda Zhang, Jiahao Li, Yunxin Li, Shijue Huang, Wanjun Zhong, Kuanye Li,
  Jiale Yang, Yu~Miao, Woyu Lin, Longxiang Liu, Xu~Jiang, Qianli Ma, Jingyu Li,
  Xiaojun Xiao, Kai Cai, Chuang Li, Yaowei Zheng, Chaolin Jin, Chen Li, Xiao
  Zhou, Minchao Wang, Haoli Chen, Zhaojian Li, Haihua Yang, Haifeng Liu, Feng
  Lin, Tao Peng, Xin Liu, and Guang Shi.
\newblock {UI-TARS: Pioneering Automated GUI Interaction with Native Agents},
  2025.
\newblock URL \url{https://arxiv.org/html/2501.12326v1}.

\bibitem[Qin et~al.(2026)Qin, Yue, Hua, Fu, and Ren]{qin2026eam}
Zerui Qin, Sheng Yue, Xingyuan Hua, Yongjian Fu, and Ju~Ren.
\newblock {Executable Agentic Memory for GUI Agent}, 2026.
\newblock URL \url{https://arxiv.org/html/2605.12294v1}.

\bibitem[Rawles et~al.(2025)Rawles, Clinckemaillie, Chang, Waltz, Lau, Fair,
  Li, Bishop, Li, Campbell-Ajala, Toyama, Berry, Tyamagundlu, Lillicrap, and
  Riva]{rawles2024androidworld}
Christopher Rawles, Sarah Clinckemaillie, Yifan Chang, Jonathan Waltz,
  Gabrielle Lau, Marybeth Fair, Alice Li, William Bishop, Wei Li, Folawiyo
  Campbell-Ajala, Daniel Toyama, Robert Berry, Divya Tyamagundlu, Timothy
  Lillicrap, and Oriana Riva.
\newblock {AndroidWorld: A Dynamic Benchmarking Environment for Autonomous
  Agents}.
\newblock In \emph{International Conference on Learning Representations}, 2025.
\newblock URL
  \url{https://proceedings.iclr.cc/paper_files/paper/2025/hash/01a83bc2f2732a58e6aa731e659e7101-Abstract-Conference.html}.

\bibitem[Song et~al.(2026)Song, Zhang, Song, and Zhang]{song2026appagentclaw}
Zhixue Song, Zhiheng Zhang, Yi~Song, and Chi Zhang.
\newblock {AppAgent-Claw: CLI Is All You Need for GUI Automation}, 2026.
\newblock URL \url{https://arxiv.org/html/2606.05171v1}.

\bibitem[Wang et~al.(2026)Wang, Wu, Zhang, Zhang, Yao, Faisal, Peng, Qin, Nath,
  Lin, Bansal, Zhang, Rajmohan, Gao, and Yao]{wang2026webxskill}
Zhaoyang Wang, Qianhui Wu, Xuchao Zhang, Chaoyun Zhang, Wenlin Yao, Fazle~Elahi
  Faisal, Baolin Peng, Si~Qin, Suman Nath, Qingwei Lin, Chetan Bansal, Dongmei
  Zhang, Saravan Rajmohan, Jianfeng Gao, and Huaxiu Yao.
\newblock {WebXSkill: Skill Learning for Autonomous Web Agents}, 2026.
\newblock URL \url{https://arxiv.org/html/2604.13318v2}.

\bibitem[Wang et~al.(2025{\natexlab{a}})Wang, Xu, Wang, Zhang, Yan, Zhang,
  Huang, and Ji]{wang2025mobileagente}
Zhenhailong Wang, Haiyang Xu, Junyang Wang, Xi~Zhang, Ming Yan, Ji~Zhang, Fei
  Huang, and Heng Ji.
\newblock {Mobile-Agent-E: Self-Evolving Mobile Assistant for Complex Tasks}.
\newblock In \emph{NeurIPS 2025 Workshop on Scaling Environments for Agents},
  2025{\natexlab{a}}.
\newblock URL
  \url{https://openreview.net/pdf/12455a402402e23beff4d62834e504c04fe42cda.pdf}.

\bibitem[Wang et~al.(2025{\natexlab{b}})Wang, Mao, Fried, and
  Neubig]{wang2024awm}
Zora~Zhiruo Wang, Jiayuan Mao, Daniel Fried, and Graham Neubig.
\newblock {Agent Workflow Memory}.
\newblock In \emph{Proceedings of the 42nd International Conference on Machine
  Learning}, volume 267 of \emph{Proceedings of Machine Learning Research},
  pp.\  63897--63911, 2025{\natexlab{b}}.
\newblock URL \url{https://proceedings.mlr.press/v267/wang25bx.html}.

\bibitem[Wen et~al.(2024)Wen, Li, Liu, Zhao, Yu, Li, Jiang, Liu, Zhang, and
  Liu]{wen2023autodroid}
Hao Wen, Yuanchun Li, Guohong Liu, Shanhui Zhao, Tao Yu, Toby Jia-Jun Li, Shiqi
  Jiang, Yunhao Liu, Yaqin Zhang, and Yunxin Liu.
\newblock {AutoDroid: LLM-powered Task Automation in Android}.
\newblock In \emph{Proceedings of the 30th Annual International Conference on
  Mobile Computing and Networking}, 2024.
\newblock \doi{10.1145/3636534.3649379}.
\newblock URL \url{https://yunxinliu.github.io/}.

\bibitem[Wen et~al.(2025)Wen, Tian, Pavlov, Du, Li, Chang, Zhao, Liu, Liu,
  Zhang, and Li]{wen2025autodroidv2}
Hao Wen, Shizuo Tian, Borislav Pavlov, Wenjie Du, Yixuan Li, Ge~Chang, Shanhui
  Zhao, Jiacheng Liu, Yunxin Liu, Ya-Qin Zhang, and Yuanchun Li.
\newblock {AutoDroid-V2: Boosting SLM-based GUI Agents via Code Generation}.
\newblock In \emph{Proceedings of the 23rd Annual International Conference on
  Mobile Systems, Applications and Services}, 2025.
\newblock \doi{10.1145/3711875.3729134}.
\newblock URL \url{https://arxiv.org/html/2412.18116v3}.

\bibitem[Xiao et~al.(2026)Xiao, Wang, Wang, Liu, Chai, Pan, Zhou, Chen, Wen,
  and Li]{xiao2026uimem}
Han Xiao, Guozhi Wang, Hao Wang, Shilong Liu, Yuxiang Chai, Yue Pan, Yufeng
  Zhou, Xiaoxin Chen, Yafei Wen, and Hongsheng Li.
\newblock {UI-Mem: Self-Evolving Experience Memory for Online Reinforcement
  Learning in Mobile GUI Agents}, 2026.
\newblock URL \url{https://arxiv.org/html/2602.05832v1}.

\bibitem[Yang et~al.(2025)Yang, Dou, Feng, Huang, Nguyen, You, Attia, Yang,
  Feng, Zhang, Ramrakhya, Jia, Nichols, Toshev, Yang, and
  Gan]{yang2025ferretuilite}
Zhen Yang, Zi-Yi Dou, Di~Feng, Forrest Huang, Anh Nguyen, Keen You, Omar Attia,
  Yuhao Yang, Michael Feng, Haotian Zhang, Ram Ramrakhya, Chao Jia, Jeffrey
  Nichols, Alexander Toshev, Yinfei Yang, and Zhe Gan.
\newblock {Ferret-UI Lite: Lessons from Building Small On-Device GUI Agents},
  2025.
\newblock URL \url{https://arxiv.org/html/2509.26539v1}.

\bibitem[Zeng et~al.(2026)Zeng, Hua, Zou, Cai, Feris, and
  Luo]{zeng2026mementogui}
Ziyun Zeng, Hang Hua, Bocheng Zou, Mu~Cai, Rogerio Feris, and Jiebo Luo.
\newblock {MementoGUI: Learning Agentic Multimodal Memory Control for
  Long-Horizon GUI Agents}, 2026.
\newblock URL \url{https://arxiv.org/html/2605.18652v1}.

\bibitem[Zhang et~al.(2025{\natexlab{a}})Zhang, Feng, Zhao, Zhao, Gong, Sun,
  Du, Hua, Xia, and Chen]{zhang2025mobiagent}
Cheng Zhang, Erhu Feng, Xi~Zhao, Yisheng Zhao, Wangbo Gong, Jiahui Sun, Dong
  Du, Zhichao Hua, Yubin Xia, and Haibo Chen.
\newblock {MobiAgent: A Systematic Framework for Customizable Mobile Agents},
  2025{\natexlab{a}}.
\newblock URL \url{https://arxiv.org/html/2509.00531v1}.

\bibitem[Zhang et~al.(2025{\natexlab{b}})Zhang, Yang, Liu, Li, Han, Chen,
  Huang, Fu, and Yu]{zhang2023appagent}
Chi Zhang, Zhao Yang, Jiaxuan Liu, Yanda Li, Yucheng Han, Xin Chen, Zebiao
  Huang, Bin Fu, and Gang Yu.
\newblock {AppAgent: Multimodal Agents as Smartphone Users}.
\newblock In \emph{Proceedings of the 2025 CHI Conference on Human Factors in
  Computing Systems}, 2025{\natexlab{b}}.
\newblock \doi{10.1145/3706598.3713600}.
\newblock URL \url{https://appagent-official.github.io/}.

\bibitem[Zhang et~al.(2026)Zhang, Shao, Li, Lin, Fu, Wang, Jiao, Lu, Liu,
  Zhang, and Yu]{zhang2026mmskills}
Kangning Zhang, Shuai Shao, Qingyao Li, Jianghao Lin, Lingyue Fu, Shijian Wang,
  Wenxiang Jiao, Yuan Lu, Weiwen Liu, Weinan Zhang, and Yong Yu.
\newblock {MMSkills: Towards Multimodal Skills for General Visual Agents},
  2026.
\newblock URL \url{https://arxiv.org/html/2605.13527v1}.

\bibitem[Zhao et~al.(2026{\natexlab{a}})Zhao, Shanmugham, Roy, and
  Xu]{zhao2026echopath}
Yao Zhao, Aditya Shanmugham, Swastik Roy, and Yanxun Xu.
\newblock {EchoPath: Execution-Level Replayable Memory for GUI Agents},
  2026{\natexlab{a}}.
\newblock URL \url{https://arxiv.org/html/2609.16635v1}.

\bibitem[Zhao et~al.(2026{\natexlab{b}})Zhao, Liew, Yang, Yang, Luo, Sahoo,
  Savarese, and Li]{zhao2026gpa}
Zirui Zhao, Jun~Hao Liew, Yan Yang, Wenzhuo Yang, Ziyang Luo, Doyen Sahoo,
  Silvio Savarese, and Junnan Li.
\newblock {GPA: Learning GUI Process Automation from Demonstrations},
  2026{\natexlab{b}}.
\newblock URL \url{https://arxiv.org/html/2604.01676v2}.

\bibitem[Zheng et~al.(2025)Zheng, Fatemi, Jin, Wang, Gandhi, Song, Gu,
  Srinivasa, Liu, Neubig, and Su]{zheng2025skillweaver}
Boyuan Zheng, Michael~Y. Fatemi, Xiaolong Jin, Zora~Zhiruo Wang, Apurva Gandhi,
  Yueqi Song, Yu~Gu, Jayanth Srinivasa, Gaowen Liu, Graham Neubig, and Yu~Su.
\newblock {SkillWeaver: Web Agents can Self-Improve by Discovering and Honing
  Skills}, 2025.
\newblock URL \url{https://arxiv.org/html/2504.07079v1}.

\bibitem[Zhong et~al.(2026)Zhong, Faisal, Fran{\c c}a, Leesatapornwongsa,
  Szekeres, Rong, and Nath]{zhong2026actionengine}
Hongbin Zhong, Fazle Faisal, Luis Fran{\c c}a, Tanakorn Leesatapornwongsa,
  Adriana Szekeres, Kexin Rong, and Suman Nath.
\newblock {ActionEngine: From Reactive to Programmatic GUI Agents via State
  Machine Memory}, 2026.
\newblock URL \url{https://arxiv.org/html/2602.20502v1}.

\end{thebibliography}
\bibliographystyle{iclr2027_conference}

\clearpage
\appendix
\begingroup
\renewcommand{\ttdefault}{cmtt}

\section{Implementation Details}
\label{app:implementation}

The system records local transitions, extracts a graph--text template, and checks one candidate action at a time against the live instruction and interface. This section specifies the logical record schema and control flow. The transition example explains the contract at the method level.

\subsection{Record and transition schema}

A record associates an instruction with a sequence of observed states, actions, and checked outcomes. The transition contract follows the main text:
\begin{equation}
 e=(v_i,a,v_j,z),\qquad z=(g,k,s,r,\phi^{-},\phi^{+}).
\end{equation}
Here, $v_i$ and $v_j$ describe the observable source and target states; $a$ is the action; $g$ is the intended procedure step; $k$ contains typed slots; $s$ identifies the action target; $r$ records risk; and $\phi^{-}$ and $\phi^{+}$ state the required pre- and post-action evidence. A successful episode supplies demonstration evidence; each transition also needs an identifiable local outcome. An explicitly verified prefix may be retained with its truncation boundary; a failed terminal action is not a successful demonstration.

\begin{table}[htbp]
\centering\small
\caption{Logical fields of a recorded transition and their roles in selective replay.}
\label{tab:app-schema}
\begin{tabular}{@{}>{\raggedright\arraybackslash}p{.22\linewidth}>{\raggedright\arraybackslash}p{.72\linewidth}@{}}
\toprule
Field group & Content and role \\
\midrule
Record identity & Record/episode ID, instruction, application, environment, and links to the source trajectory. These identify the execution from which an edge was extracted. \\
State evidence & Before/after screenshots, UI tree when available, foreground application/window, and observable state attributes needed by the contract. Missing metadata is distinguished from a negative observation. \\
Action and selector & Action type and payload, semantic target description, and available selector evidence, such as text, role, resource identifier, or bounds. Historical bounds alone do not establish a live match. \\
Procedure and slots & Procedure intent $g$; typed slots $k$ with recorded values, live binding sources, and constraints. Values and entity identities remain explicit. \\
Checks and outcome & Preconditions $\phi^{-}$; expected effect $\phi^{+}$; observed check result; progress/completion evidence; and any truncation or failure reason. \\
Risk and provenance & Risk $r$, source record/step IDs, and links to evidence supporting admission or rejection. \\
\bottomrule
\end{tabular}
\end{table}

\subsection{Template extraction, merging, and conflict branches}

Extraction preserves stable navigation and procedure structure while exposing changeable quantities, times, and entities as typed slots. Two observations can support a shared template edge only when their intended progress, relevant source-state conditions, action semantics, and checked outcomes are compatible. Visual resemblance alone does not justify merging. For example, settings pages with a toggle off or on may share navigation but require different action preconditions.

Conflicting conditions remain separate branches. An edge that opens a product's specification sheet cannot be replaced by one that directly adds the product to the cart unless the different preconditions and effects are represented. Source-record links preserve the evidence behind each branch. A transition with missing evidence, an ungroundable target, or a failed terminal outcome does not become an executable template edge merely because another record contains a similar instruction.

\subsection{Retrieval, slot binding, and live target grounding}

Retrieval first identifies a procedure compatible with the live instruction and application context. The controller then checks candidate transitions against procedure progress and the current UI. Retrieval supplies candidates; a high similarity score does not itself authorize an action.

A slot binding connects a value in the live instruction to the action that uses it. A changed duration, for example, must update the duration-bearing action and the expected outcome. Navigation edges may remain reusable. Entity substitutions additionally require identifying the requested live product, contact, or location. An ambiguous reference does not justify silently importing a historical value.

Target grounding uses current observations to identify the control that implements the candidate action. A displaced control may remain usable if its semantics and state are supported. A missing control, conflicting identity, or blocking window instead requires rejection or fallback. The paraphrase comparison separately examines app-level semantic retrieval and soft task matching; the cumulative configurations are summarized in Section~\ref{app:configuration-matrix}.

\subsection{Precheck, postcheck, and fallback}

Before execution, the controller refreshes the evidence needed by the selected edge and checks whether the goal is already satisfied. It emits \textsc{reuse} for a supported unchanged action, \textsc{rewrite} for a supported action whose typed payload is rebound, and \textsc{reject} when evidence is missing or contradictory. Rejection executes no GUI action; fallback follows it, not as a fourth decision.

An admitted action executes once. Its postcheck verifies the expected semantic effect; a page change alone does not prove that a requested value or entity is correct. A failed postcheck blocks that edge for the current episode and returns control to the base agent. The next decision uses a fresh observation.

\begin{figure}[htbp]
\centering
\fbox{\begin{minipage}{.94\linewidth}
\small
\textbf{Algorithm 1: Guarded transition execution (method-level specification)}
\begin{enumerate}\setlength{\itemsep}{2pt}\setlength{\parskip}{0pt}
\item Input: live instruction $q$, template memory $\mathcal{M}$, base policy $\pi$, and episode budget. Initialize the episode's blocked-edge set $B=\emptyset$.
\item Observe the live UI. If supported completion evidence satisfies $q$, terminate successfully without an additional GUI action.
\item Retrieve a compatible candidate $e\notin B$; bind its slots from $q$ and ground its target in the live UI.
\item Refresh required pre-action evidence. If the goal is now satisfied, terminate. If no candidate is admissible, record \textsc{reject} (or no candidate) and invoke $\pi$ from the current state.
\item Otherwise, emit \textsc{reuse} or \textsc{rewrite}, execute the admitted action once, and re-observe the UI.
\item Evaluate $\phi^{+}$. On failure, add $e$ to $B$ and return control to $\pi$. On success, update the observed procedure progress.
\item Re-observe after fallback as well. Continue from step 2 until verified completion, a terminal failure, or budget exhaustion.
\end{enumerate}
\end{minipage}}
\caption{Control flow for one-transition admission and recovery. Human safety termination in the real-device protocol is an external evaluation event, not an additional planning step.}
\label{alg:app-replay}
\end{figure}

\paragraph{Complete transition example: changing a timer duration.}
The record instruction is ``Set a 10-minute timer'' and the live instruction is ``Set a 15-minute timer.'' Consider a duration-entry edge in a timer screen's editable field.

\begin{table}[htbp]
\centering\small
\caption{The transition contract specifies how to rewrite a timer duration from 10 to 15 minutes using the live instruction, with explicit grounding, preconditions, postconditions, and fallback.}
\label{tab:app-transition-example}
\begin{tabular}{@{}>{\raggedright\arraybackslash}p{.22\linewidth}>{\raggedright\arraybackslash}p{.72\linewidth}@{}}
\toprule
Contract field & Example value \\
\midrule
Provenance & Illustrative record \texttt{timer-record}, duration-entry step. \\
$v_i$, $v_j$ & Timer editor before entry; timer editor displaying the accepted duration after entry. \\
$g$, $a$, $s$ & Set duration; enter a duration value; live editable duration control identified from its label/role and current state. \\
$k$ & \texttt{duration}: positive duration; recorded value 600 seconds; live value 900 seconds, explicitly bound from the instruction. \\
$r$ & Reversible local edit; this edge does not start the timer or perform an external transaction. \\
$\phi^{-}$ & Correct timer editor is active, no blocking window is present, and the duration control is editable and unambiguously grounded. \\
Decision and payload & \textsc{rewrite}; render 900 seconds in the control's accepted format. Navigation into the editor can still be reused. \\
$\phi^{+}$ & The displayed duration denotes 15 minutes. A changed screen or an unaccepted text entry is insufficient. \\
Failure and completion & Reject if preconditions fail; block and fall back if the postcheck fails. Starting the timer is a separate edge, whose completion check requires a running timer consistent with the requested duration. \\
\bottomrule
\end{tabular}
\end{table}

\section{Experimental Setup and Evaluation Protocol}
\label{app:experimental-setup}

\subsection{AndroidWorld and base agents}

AndroidWorld provides parameterized task initialization and programmatic completion checks. Its original suite contains 116 tasks across 20 applications; these numbers describe the benchmark rather than the evaluated subset for an individual model. Runs use English interfaces and instructions. Each base agent's successful LLM-only executions supply the records for its template graphs, and assisted execution uses that agent as fallback. Record and live tasks are similar, with some pairs differing in task details or parameter values. The AndroidWorld success rate in Table~\ref{tab:overall-results} is computed as base success multiplied by the success rate conditional on records from successful base executions.

The five base agents are GUI-Owl-7B, MAI-UI-8B, GELab-Zero-4B-preview, AutoGLM-Phone-9B-Multilingual, and UI-TARS-1.5-7B.

\subsection{Physical devices and language settings}

Real-device testing uses OnePlus 13, OPPO Find X8, Redmi K50, and Huawei Pura 80 phones, spanning Android and HarmonyOS. The real-device study comprises 100--110 tasks. Tasks that fail on their initial run are retested to confirm that the failures are not caused by non-technical factors. Table~\ref{tab:app-tested-apps} lists applications included in testing. The inventory describes application coverage and does not imply that every application was evaluated on every device or with every model.

Real-device testing uses multiple languages, including Chinese interfaces and Chinese task instructions. The paper presents task descriptions in English for readability. This presentation language does not imply that execution used English or that translation preceded execution. Section~\ref{app:example-presentation} explains the use of neutral aliases in the examples.

\subsection{Record--live comparisons and outcome judgment}

A record--live pair connects an earlier verified execution with a current request and device state. In the three quantitative conditions, background-state changes preserve the instruction while changing the current state, parameter changes modify a requested value, and paraphrases change wording while preserving intent. Within a comparison, the four configurations receive the same live instruction, initial state, and completion criterion. The real-device ablations use tasks designed to represent practical usage scenarios and executed on physical phones. The same designed task set is used across configurations, without selecting tasks on the basis of prior LLM-only success. Applications and representative task descriptions are provided in Appendix~\ref{app:variation-cases}.

During real-device execution, an observer beside the phone judges whether the resulting object, value, and final state satisfy the live request. The observer evaluates outcomes without approving plans, correcting slot bindings, or selecting actions during execution. The completion boundary follows the requested task: adding an item to a cart or reaching an order page can be a terminal goal without completing a purchase. If execution crosses such a boundary and reaches a payment-password entry page, the observer terminates the process and marks the episode as a failure. This failure remains in the evaluated denominator. The same rule applies across compared configurations.

\subsection{Cumulative component configurations}
\label{app:configuration-matrix}

Table~\ref{tab:app-component-matrix} expands the configuration names used in the main-text ablations. Each row is a separate comparison, and additions accumulate within that row. Replay-only execution uses a stored procedure from an available graph; it is distinct from the LLM-only execution used as the efficiency reference.

\begin{table}[htbp]
\centering\small
\caption{Cumulative component configurations. C0 uses replay only; C1 and C2 add the listed mechanisms without model fallback, and C2 retains the C1 addition. C3 uses the full controller with base-model fallback and includes semantic routing in the paraphrase comparison.}
\label{tab:app-component-matrix}
\setlength{\tabcolsep}{4pt}
\renewcommand{\arraystretch}{1.16}
\begin{tabular}{@{}>{\raggedright\arraybackslash}p{.19\linewidth}>{\raggedright\arraybackslash}p{.32\linewidth}>{\raggedright\arraybackslash}p{.41\linewidth}@{}}
\toprule
Condition & C1: first addition & C2: additional component \\
\midrule
Background state & State recheck / TOCTOU-aware replay & Goal preservation: detect an already satisfied goal. \\
Parameter change & Parameter generalization & Instruction--outcome agreement: check the requested value. \\
Paraphrase & App-level semantic retrieval & Soft task matching: associate wording with a stored procedure. \\
\bottomrule
\end{tabular}
\end{table}

C0 executes the stored procedure through replay only. C1 and C2 augment replay with the corresponding cumulative mechanisms in Table~\ref{tab:app-component-matrix}, without model fallback. Only the full configuration C3 enables fallback to the base model.
\FloatBarrier

\section{Real-World Applications and Task Examples}
\label{app:variation-cases}

\subsection{Applications included in testing}

Table~\ref{tab:app-tested-apps} summarizes the applications used in real-world testing. The coverage includes communication, media, information access, shopping, travel, and productivity. AI-assistant applications in this inventory are target interfaces operated by the GUI agent; they are separate from the five base agents evaluated in the main text. Ctrip is included because it appears in the tested hotel-search tasks.

\begin{table}[htbp]
\centering\small
\caption{Applications included in real-world testing, grouped by usage category. Names are given in English or romanized product names. This inventory documents coverage rather than per-application success rates or a balanced application-level benchmark.}
\label{tab:app-tested-apps}
\renewcommand{\arraystretch}{1.13}
\begin{tabular}{@{}>{\raggedright\arraybackslash}p{.25\linewidth}>{\raggedright\arraybackslash}p{.71\linewidth}@{}}
\toprule
Category & Applications \\
\midrule
Social / messaging & WeChat, QQ, Feishu \\
Short video / video & Douyin, Kuaishou, Bilibili, YouTube, Tencent Video, iQIYI, Youku \\
Communities / content & Xiaohongshu (RedNote), Weibo, Zhihu, Douban, Baidu Tieba, Reddit, X, Instagram \\
Music / audio & QQ Music, NetEase Cloud Music, Kugou Music \\
News / reading & Toutiao, Tencent News, WeRead \\
Browsers / search & Google Chrome, Microsoft Edge, Quark, Baidu, Google, Bing \\
Maps / mobility & Amap, Baidu Maps, Google Maps, DiDi, Uber \\
Food / local services & Meituan, Dianping, Taobao Instant Commerce (Taobao Shangou; formerly Ele.me) \\
E-commerce / shopping & Taobao, JD.com, Pinduoduo, Xianyu (Idle Fish), Amazon \\
Travel / hotels & Ctrip \\
Email / calendars & Gmail, Outlook, QQ Mail, NetEase Mail \\
Cloud storage / files & Baidu Netdisk, Alibaba Cloud Drive, Quark Cloud Drive \\
AI assistants & ChatGPT, Claude, Gemini, DeepSeek, Doubao, Kimi, Qwen (Tongyi Qianwen), Tencent Yuanbao, ERNIE Bot, Perplexity, Grok \\
\bottomrule
\end{tabular}
\end{table}
\FloatBarrier

\subsection{Representative task descriptions}
\label{app:example-presentation}

Some task instructions contain account identifiers, travel details, or personal shopping and content preferences. Table~\ref{tab:app-task-examples} presents shortened English descriptions of tested tasks using neutral aliases for creators, stores, products, and destinations. A demographic-specific clothing category is described as the requested category, retaining its role as a selection constraint. Exact travel dates are represented by check-in and check-out slots. These presentation edits retain the relevant task structure while reducing unnecessary identifying and preference-specific detail.

\begin{table}[htbp]
\centering\small
\caption{Neutralized task descriptions drawn from real-device testing. Aliases replace identifying or preference-specific details; rows describe tasks rather than their execution outcomes.}
\label{tab:app-task-examples}
\renewcommand{\arraystretch}{1.16}
\begin{tabular}{@{}>{\raggedright\arraybackslash}p{.23\linewidth}>{\raggedright\arraybackslash}p{.73\linewidth}@{}}
\toprule
Application & English task description \\
\midrule
Taobao & Add the specified smartphone, Smartphone A, to the cart. \\
Taobao & Find the best-selling jacket in the requested category priced between CNY 100 and CNY 200 for purchase. \\
Ctrip & Check hotel prices in Destination A for a two-night stay with the specified check-in and check-out dates. \\
Taobao Instant Commerce & Find a kumquat--lemon drink and add one cup to the cart. \\
Xiaohongshu & View the latest post by Creator A. \\
Taobao & Search for Store A, open the first matching store, search within it for Product B, select the default configuration, and add the product to the cart. \\
\bottomrule
\end{tabular}
\end{table}
\FloatBarrier

The examples cover direct search, constrained product selection, date-conditioned information retrieval, creator lookup, and search within a store. A phrase such as ``first result'' is a selection instruction in the current interface; it is not a persistent identifier for a historical item. Likewise, ``default configuration'' refers to the selected live product's options. Reuse must check the intended product or store and its current state before executing a stored action.

\subsection{Shared suffixes after different navigation paths}
\label{app:trajectories}

The following task pairs illustrate how different prefixes can reach a compatible state from which later transitions may be reused. The discussion explains candidate reuse from the task descriptions rather than reporting step-level execution outcomes. A shared suffix remains a sequence of individually checked transitions.

\paragraph{Taobao: search entry versus cart entry.}
\begin{description}\setlength{\itemsep}{2pt}
\item[Record task.] Open Taobao, search for Smartphone A, open the first matching product, tap Add to Cart, select the default configuration, and confirm adding one unit to the cart.
\item[Live task.] Open the cart, open its first item corresponding to Smartphone A, and add one more unit of that product to the cart.
\item[Shared suffix.] Once product identity and the required option-selection state agree, the relevant product-configuration and add-to-cart transitions can be considered for reuse. Navigation through search is unnecessary for the cart-entry task.
\end{description}

The live request explicitly asks for an additional unit. An existing cart entry therefore does not by itself satisfy the goal: the expected outcome is an increase by one in the requested product/configuration quantity. The cart's first item must be checked against Smartphone A, and a changed selection state can invalidate an individual suffix transition. The instruction's extra confirmation tap is a UI step, not a request to add another unit.

\paragraph{Zhihu: search entry versus saved-answer entry.}
\begin{description}\setlength{\itemsep}{2pt}
\item[Record task.] Search Zhihu for ``large language model inference,'' open the first answer, like and save it, then open the author's profile and follow the author.
\item[Live task.] Open the saved collection, open the first answer corresponding to ``large language model inference,'' then open the author's profile and follow the author.
\item[Shared suffix.] From a compatible answer page, opening the corresponding author profile and following that author forms the common suffix. Search, liking, and saving belong to the earlier path and are not required by the live instruction.
\end{description}

The live answer must determine the author identity; search rank or saved-item position alone cannot establish that it is the historical author. The current follow state must also be checked so that an already satisfied goal does not trigger an inverse action.

\subsection{Relation to the variation taxonomy}

The main-text quantitative comparisons cover background-state changes, parameter changes, and paraphrased instructions. The examples below explain these conditions. They are illustrative controlled pairs rather than additional measured episodes.

\paragraph{Background-state changes.}
\label{app:background-state}
For ``Turn on dark mode,'' the record begins with dark mode off while the live state already has it on. The instruction and relevant page remain fixed; the changed factor is the toggle state. Correct completion preserves the enabled state without another toggle. This differs from the Taobao suffix example, whose live instruction requests an additional item.

\paragraph{Parameter changes.}
\label{app:parameter-change}
A record sets a 10-minute timer, whereas the live instruction requests a 15-minute timer from the same initial timer state. The duration changes while the procedure remains stable. Navigation can be reused, but duration entry and its expected effect use the live value. Table~\ref{tab:app-transition-example} specifies this illustrative transition. Shopping quantities and hotel dates provide other typed values in the task descriptions; their presence alone does not establish a controlled parameter-change experiment.

\paragraph{Paraphrased instructions.}
\label{app:paraphrased-instructions}
``Turn on Wi-Fi'' and ``Enable Wi-Fi'' express the same goal from the same initial off state. The changed factor is wording, so retrieval should identify the same procedure while execution still checks the live toggle. A contextual reference such as ``my usual item'' instead requires additional evidence and is not an equivalent paraphrase when its referent is unresolved.

\paragraph{Other variation categories.}
Layout/workflow changes, modal interruptions, and entity substitutions complete the six-category taxonomy. A displaced product requires live grounding; an inserted selection page changes the procedure; a blocking dialog changes action admissibility; and a substituted destination requires resolving the new place. These categories explain possible mismatches but are separate from the three quantitative conditions. The shared-suffix examples likewise explain task structure without adding a new quantitative evaluation.
\FloatBarrier

\section{Metric Definitions and Supplementary Results}
\label{app:metrics-results}

\subsection{Task success and evaluation scope}

Task success is the number of verified successful episodes $n$ divided by the number of evaluated eligible episodes $N$. Failed actions, budget exhaustion, and human safety terminations contribute failures. Setup exclusions, if any, are distinct from failures within the evaluated set. For AndroidWorld record-conditional success, $N$ counts trials supported by records from successful base-agent executions, and $n$ counts successful trials within that set. Table~\ref{tab:overall-results} reports base success multiplied by this conditional rate, rounded to one decimal place. This product is computed from the reported aggregate percentages rather than a separate full-suite evaluation. In the real-world ablations, all configurations are evaluated on the same tasks designed for practical usage scenarios and executed on physical phones. This task set is not restricted to tasks previously completed successfully by the base model. The full configuration combines replay with model fallback; its success rate is not conditional on replay-only success. The ablation plots report these per-configuration success rates directly. The real-world ``Success score with replay'' column in Table~\ref{tab:overall-results} is the arithmetic product of the replay-only and full-configuration percentages, divided by 100; it is a derived score, not the measured success rate of the full controller. Real-device results identify the variation and configuration. Counts are not reconstructed from rounded percentages.

\subsection{Executed reuse and zero-action completion}

Executed reuse counts actions supplied by replay without a model call. A local rewrite counts only when it meets this condition. Retrieved or rejected candidates are not executed reuse. Total executed actions include replay, fallback, and recovery actions. All reported replay rates use the task-mean definition $r_{\mathrm{task}}$ in Equation~\ref{eq:replay-rates}: first compute the ratio of replayed actions to all executed actions for each episode with at least one action, then take the arithmetic mean across episodes. Each episode receives equal weight. The pooled step rate $r_{\mathrm{step}}$ instead weights episodes by their action counts; it is used in the explanatory cost relationships below, not as the reported replay-rate aggregation.

If the goal is already satisfied and the controller correctly stops without a GUI action, the episode remains eligible for task success. Such an episode is a zero-action completion: its action-based reuse ratio is undefined, so it is excluded from action-based reuse-rate aggregation. A false completion declaration is a failure. Any model calls used to verify completion still contribute to cost.

\subsection{Paired costs, coverage, and timing boundaries}

An efficiency pair matches the task instance, initial state, model, and completion criterion between an LLM-only execution and execution assisted by the full replay controller. The real-device replay-only success baseline is not this cost reference. Cost and execution-efficiency statistics include only pairs in which both executions succeed. Task success is evaluated over the eligible task set and reported separately from efficiency on successful pairs. Paired-success coverage is defined as the number of successful pairs divided by all eligible pairs.

Execution time covers the complete online task execution, from its start to task completion, including UI acquisition, model inference, device actions, waits for page or state changes, retrieval and binding, checks, and recovery. External setup/reset and record/template construction are separate costs. For successful pair $i$, let $T_{0,i}$ be the LLM-only execution time and $T_{1,i}$ the full-controller execution time. The reported speedup is the arithmetic mean of the per-task time ratios:
\begin{equation*}
 \overline{S}=\frac{1}{|\mathcal{P}|}\sum_{i\in\mathcal{P}}\frac{T_{0,i}}{T_{1,i}},
\end{equation*}
where $\mathcal{P}$ is the set of pairs in which both executions succeed. This gives each pair equal weight; it is not the ratio of total baseline time to total assisted time. The cost relationships below explain the metrics without replacing the measured task-level aggregation.

\subsection{Derived cost relationships}

The following relationships distinguish replay rate, model-call savings, and runtime savings. They explain the metrics but are not used to compute the experimental values. Let $N_0$ and $N_1$ be pooled GUI-action counts on the same paired tasks without and with replay, $R$ the number of model-free replay actions, and $r=R/N_1$. With one model call per baseline action and per non-replayed assisted action, plus $C_{\mathrm{extra}}$ additional assisted calls,
\begin{equation}
 C_0=N_0,\qquad C_1=N_1-R+C_{\mathrm{extra}},\qquad
 d_{\mathrm{LLM}}=1-\frac{C_1}{C_0}.
\end{equation}
Additional calls include model-based retrieval, checking, and recovery not already counted. Token reduction is $d_{\mathrm{tok}}=1-K_1/K_0$, where $K_0,K_1$ include all input/output and visual-token charges. Physical action reduction, $1-N_1/N_0$, differs because reused actions still execute on the device.

If $N_1=N_0$, $C_{\mathrm{extra}}=0$, and the average remaining call uses $\alpha$ times the baseline tokens, then
\begin{equation}
 d_{\mathrm{LLM}}=r,\qquad d_{\mathrm{tok}}\approx 1-\alpha(1-r).
\end{equation}
These assumptions need not hold in the reported runs, so replay rate is not substituted for LLM-step or token reduction. No fixed value of $\alpha$ is used to derive the experimental results. Under equal model-call latency $\tau$, no additional calls, and replay overhead $\epsilon$ per replayed action,
\begin{equation}
 S_{\mathrm{decision}}=\frac{N_0\tau}{(N_1-R)\tau+R\epsilon}
 \ \longrightarrow\ \frac{1}{1-r}
 \quad\text{when }N_1=N_0\text{ and }\epsilon/\tau\to0.
\end{equation}
This limit excludes device execution and waits, additional model calls, and construction costs. For each successful task pair, end-to-end speedup is $T_{0,i}/T_{1,i}$ with matched timing boundaries; the reported speedup averages these ratios. Reuse alone cannot determine it.

\subsection{Execution efficiency}

Table~\ref{tab:app-efficiency-summary} reports efficiency ranges across five base agents per condition. Speedups are arithmetic means of per-task execution-time ratios over successful pairs, using LLM-only execution as the cost reference.

\begin{table}[htbp]
\centering\small
\caption{Execution-efficiency ranges across five base agents. Replay, token reduction, and LLM-step reduction are percentages; speedup is a multiplier.}
\label{tab:app-efficiency-summary}
\setlength{\tabcolsep}{4pt}
\begin{tabular*}{\linewidth}{@{\extracolsep{\fill}}lcccc@{}}
\toprule
Condition & Replay & Token $\downarrow$ & LLM steps $\downarrow$ & Speedup \\
\midrule
AndroidWorld & 82.0--87.5 & 78.0--89.6 & 71.1--79.8 & 1.2--1.9$\times$ \\
Background-state changes & 72.6--78.0 & 65.4--78.9 & 80.6--88.8 & 1.9--3.6$\times$ \\
Parameter changes & 70.7--79.8 & 68.3--78.7 & 73.1--85.8 & 1.8--3.5$\times$ \\
Paraphrased instructions & 73.2--79.2 & 60.8--79.6 & 76.0--86.8 & 2.9--3.9$\times$ \\
\bottomrule
\end{tabular*}
\end{table}

AndroidWorld repeatedly acquires UI structures before and after actions, reducing the share of execution time spent on the LLM inference that replay can avoid. In the real-device implementation, we optimize UI-structure acquisition and reduce its frequency, increasing the fraction of runtime that benefits from inference savings. This distinction explains why large LLM-step reductions do not translate directly into equally large execution-time speedups. These ratios use the complete execution-time boundary and successful-pair aggregation defined above.

\FloatBarrier

\section{Task Example Presentation and Data Privacy}
\label{app:reproducibility}

\paragraph{Privacy and neutral presentation.}
The task descriptions use stable aliases within each example so that product, store, creator, and destination relationships remain understandable. App names identify the tested interfaces; personal identifiers and preference-specific details are unnecessary for explaining the reuse mechanism. Numeric task constraints and distinctions such as ``one unit'' versus ``one additional unit'' remain explicit. English translation, anonymization, and neutral wording are presentation choices; they do not change reported aggregate results or substitute for a fairness evaluation.

Anonymization applies to the textual examples in this appendix. Sharing screenshots, UI trees, action payloads, or logs requires separate review for account identifiers, names, contact details, addresses, private messages, and payment information. Masking must preserve the identity and value relationships needed to interpret a decision.
\endgroup

\end{document}